\documentclass[letterpaper, fontsize=10.5pt]{scrartcl}	
\usepackage[margin=0.6in]{geometry}
\usepackage[T1]{fontenc}
\usepackage{fourier}
\usepackage[english]{babel}				\usepackage[normalem]{ulem}											
\usepackage[protrusion=true,expansion=true]{microtype}				
\usepackage{amsmath,amsfonts,amsthm}	

\usepackage{multicol}									
\usepackage[pdftex]{graphicx}														
\usepackage{url}
\usepackage{wrapfig}
\usepackage{subcaption}
\usepackage{booktabs}
\usepackage{siunitx}
\usepackage{enumitem}

\usepackage{pdfpages}
\usepackage{fancyhdr}
\usepackage[numbers,sort&compress]{natbib}

\newcommand{\jkfont}{
  \bfseries
  \color{orange}
}
\DeclareTextFontCommand{\jk}{\jkfont}
\newcommand{\gskfont}{
  \bfseries
  \color{red}
}
\DeclareTextFontCommand{\gsk}{\gskfont}
\newcommand{\editfont}{
  \bfseries
  \color{green}
}
\DeclareTextFontCommand{\edit}{\editfont}

\usepackage{titlesec, blindtext, color}
	\definecolor{light-blue}{rgb}{0.8,0.85,1}
	\definecolor{airforceblue}{rgb}{0.36, 0.54, 0.66}
	\definecolor{azure}{rgb}{0.0, 0.5, 1.0}
	\definecolor{bleudefrance}{rgb}{0.19, 0.55, 0.91}
	\definecolor{blue(munsell)}{rgb}{0.0, 0.5, 0.69}
	\definecolor{darkmidnightblue}{rgb}{0.0, 0.2, 0.4}
	\definecolor{steelblue}{rgb}{0.27, 0.51, 0.71}
	\definecolor{tealblue}{rgb}{0.21, 0.46, 0.53}
	\definecolor{yaleblue}{rgb}{0.06, 0.3, 0.57}
	\definecolor{applered}{rgb}{0.89, 0.02, 0.17}
	\definecolor{brickred}{rgb}{0.8, 0.25, 0.33}

\usepackage{sectsty}												
\allsectionsfont{\normalfont \large \scshape \color{yaleblue} \bfseries}	
\usepackage{setspace}

\usepackage[pdftex,unicode, implicit]{hyperref}
\hypersetup{
    unicode=true,          
    linktocpage=true,
    pdftoolbar=true,        
    pdfmenubar=true,        
    pdffitwindow=true,     
    pdfstartview={FitH},    
    pdfnewwindow=true,      
    colorlinks=true,       
    linkcolor=yaleblue,          
    citecolor=yaleblue,        
    filecolor=yaleblue,      
    urlcolor=yaleblue, 
   dvipdfm=true
}

\newcommand{\horrule}[1]{\rule{\linewidth}{#1}} 	

\title{         \normalfont 							
		\vspace{-1.025in} 	
		{\color{yaleblue}\horrule{2pt}} \\
		\Large	
		{\color{yaleblue}{\textsc{{\textbf{Solar Flare Energy Partitioning and Transport - The Gradual Phase}}}}}\\ 
	    \small
		\textbf{Graham S. Kerr$^{1,2}$, Meriem Alaoui$^{1,2}$, Joel C. Allred$^{2}$, Nicolas H. Bian$^{3}$, Brian R. Dennis$^{2}$, A. Gordon Emslie$^{3}$, Lyndsay Fletcher$^{4,5}$, Silvina Guidoni$^{6}$, Laura A. Hayes$^{2}$, Gordon D. Holman$^{2~(\mathrm{Emeritus})}$, Hugh S. Hudson$^{4}$, Judith T. Karpen$^{2}$, Adam F. Kowalski$^{7,8}$,\\Ryan O. Milligan$^{9}$, Vanessa Polito$^{10}$, Jiong Qiu$^{11}$, Daniel F. Ryan$^{6,2}$}\\
		\small
		\textsl{(1) Catholic U. of America, (2) NASA/GSFC, (3) Western Kentucky U., (4) U. of Glasgow, (5) U. of Oslo (6) American U., (7) U. of Colorado at Boulder, (8) National Solar Obs., (9) Queens U. Belfast, (10) Bay Area Environmental Research Inst., (11) Montana State U.}\\
		\small
		{\color{yaleblue}White Paper in response to NASA's \textsl{Heliophysics 2050 Workshop} Solicitation}\\
		\vspace{-0.05in}
		{\color{yaleblue}\horrule{2pt}}
		\vspace{-.25in} 
}

\date{}

\begin{document}
\begin{spacing}{1}
\maketitle
\vspace{-1in}


\end{spacing}
\thispagestyle{empty}

\vspace{-0.1in}
\section{Overview}
\vspace{-0.15in}

Solar flares are a fundamental component of solar eruptive events (SEEs; along with solar energetic particles, SEPs, and coronal mass ejections, CMEs). Flares are the first component of the SEE to impact our atmosphere, which can set the stage for the arrival of the associated SEPs and CME. Magnetic reconnection drives SEEs by restructuring the solar coronal magnetic field, liberating a tremendous amount of energy which is partitioned into various physical manifestations: particle acceleration, mass and magnetic-field eruption, atmospheric heating, and the subsequent emission of radiation as solar flares. To explain and ultimately predict these geoeffective events, the heliophysics community requires a comprehensive understanding of the processes that transform and distribute stored magnetic energy into other forms, including the broadband radiative enhancement that characterises flares.

This white paper discusses energy transport during the gradual phase of flares.  We discuss the impulsive phase in a related white paper. The impulsive phase is the initial dramatic release and transport of flare energy, typically associated with the presence of hard X-ray (HXR) observations. HXR emission is the signature of accelerated nonthermal electrons, produced mainly from their bombardment of the dense chromosphere. The precipitation of these accelerated nonthermal electrons is thought to be the primary vehicle by which flare energy is transported from the coronal release site throughout the Sun's atmosphere. This is followed by the longer duration gradual phase, during which flare emission decays and there is a lack of HXR emission (i.e. nonthermal electron bombardment of the lower atmosphere has ceased). There may also be significant energy release and deposition during the gradual phase, but not transported via the nonthermal particles that we typically associate with flares. We address this gradual phase here. 

While much of flare research concentrates on the impulsive phase, the gradual phase is also a topic of hot debate and scientific intrigue as it exhibits some serious discrepancies between models/theory and observations. Namely, the modelled (using radiation hydrodynamic codes) cooling times during the gradual phase of flares are significantly shorter than flare observations suggest. We must ask then, what sustains the observed gradual phase, and what physical processes are missing from our models?
 
Flare (radiation-) hydrodynamic simulations are usually driven by an intense, impulsive energy injection into an individual flux tube, the magnitudes and timescales of which are constrained by observations of, e.g., HXRs or optical/UV lightcurves.  High spatiotemporal observations of optical/UV chromospheric flare sources indicate that this injection time is short, on the order of $10-20$~s \citep[e.g.,][]{2015ApJ...807L..22G,2020ApJ...895....6G}. Following cessation of this energy input, the atmosphere undergoes rapid cooling due to large conductive heat fluxes exiting the flare-heated plasma  \citep[e.g.,][]{2020ApJ...900...18K, 2018ApJ...856..149R, 2020ApJ...895...30R}. This occurs for both individual flare sources (that is, footpoints, or individual loops) and the global flare arcade (Sun-as-a-star or summed over sub-regions). In contrast, during the observed flare gradual phase the coronal flare emission can be elevated for several tens of minutes to  hours \citep{2013ApJ...765...81R,  2016ApJ...820...14Q, 2018ApJ...856...27Z, 2012ApJS..202...11R,
2013ApJ...778...68R}, compared to only a few minutes in simulations. In the dense chromosphere, observed UV and optical sources have cooling timescales on the order of several to tens of minutes \citep{2012ApJ...744...48C, 2010ApJ...725..319Q, 2020ApJ...895....6G, 2016ApJ...833...50K}, compared to only several tens to $\sim$hundred seconds in simulations. Likely related, the duration of chromospheric ablations and condensations exhibit a similar discrepancy, with observed mass flows (inferred from Doppler shifts of spectral lines) persisting significantly longer than models can reproduce \citep{2018ApJ...856..149R, 2020ApJ...900...18K, 2020ApJ...895....6G}. Since both individual sources and the global flare exhibit longer-than-predicted cooling timescales, the resolution is unlikely to just be the activation of new loops along the flare arcade \citep[e.g.,][]{1979ApJ...229..772D}. It has been suggested that {\color{yaleblue}\textsl{energy release continues in the gradual phase, with magnitude comparable to that deposited in the impulsive phase}}. However, the origin and transport mechanism of this post-impulsive phase energy is largely unknown. Towards 2050, we must aim to determine what sustains the flare gradual phase and what ingredients are missing in flare models. Specifically, efforts should aim \\

\vspace{-0.15in}
 \textbf{{\color{yaleblue}(1)} To determine the magnitude, source, and transport mechanisms of energy deposition during the gradual phase}; \\

\vspace{-0.15in}
 \textbf{{\color{yaleblue}(2)} To improve our modelling and observational constraints of the gradual phase, and our understanding of important plasma physical processes such as turbulence and non-local effects (which have far reaching effects beyond flares)}.

\vspace{-0.2in}
\section{Post-Impulsive Phase Heating}
\vspace{-0.15in}
Energy deposition and associated plasma heating during the gradual phase could be a significant fraction of the total energy released during flares, yet the character of this post-impulsive phase heating is poorly understood. Not accounting for this is a serious deficiency with our understanding of SEEs, so a concerted effort must be made to determine how much energy is required to sustain the gradual phase, and to identify the responsible mechanisms. This can be achieved through a combination of observations and modelling.

Recent work combining UV and EUV images of C class flares and 0D (loop averaged) \texttt{EBTEL} \citep{2008ApJ...682.1351K,2012ApJ...752..161C} modelling have shown interesting results \citep[][]{2016ApJ...820...14Q,2018ApJ...856...27Z}. Those studies modelled individual flare loops as a bundle of multiple threads, each energized with an intense impulsive heating episode, followed by a slow and weaker heating tail lasting several minutes. To account for duration of EUV observations, $60\%$ of flare energy had to be deposited in the short impulsive phase, with the remaining $40\%$ deposited over a span of approximately 20 minutes. 

Multi-threaded modelling of flare arcades \citep[][]{2017ApJ...851....4R} found that including effects of loop length and ribbon separation could recover some observed flare properties, including the duration of the gradual soft X-ray (SXR) emission.
Follow-up experiments, using \texttt{EBTEL}, which modelled energy deposition onto successive loops through the gradual phase with timescales guided by quasi-periodic pulsations (QPPs, see below), also successfully reproduced flare lifetimes and irradiance \citep{2020ApJ...895...30R}. In that work the heating rate was agnostic of the energy-transport mechanism (i.e., electron beams were not modelled), with no distinction between impulsive and gradual phase energy-transport mechanisms. Since we know that hard X-rays cease after the impulsive phase, we are still left with the question of what is depositing energy onto those new threads/loops during the gradual phase. If more reconnection occurs, into what forms is that energy converted and transported? 

Progress on this topic demands an accurate measurement of the energetic requirements with models capable of simulating the injection and transport of flare energy via nonthermal particles and the subsequent radiative hydrodynamic evolution of the atmosphere, including non-local and turbulent suppression of conduction (see below). Moving away from the 0D approach will allow the spatial distribution of heating to be studied, with more realistic radiative and conductive cooling. The character (magnitude and temporal profile) of post-impulsive phase energy deposition can be ascertained from comparing those models with observations. 

Reconnection Driven Current Filamentation \citep[RDCF;][]{1996ApJ...460L..73K} is a potential process for prolonged heating in flare arcades. RDCF occurs when multiple layers of steep magnetic-field gradients (current sheets) form between post-reconnection flux tubes, e.g., when plasmoids merge with the top of the flare arcade, or when the reconnection sites are distributed at random heights in the flare current sheet. This process could slowly heat the arcade into the gradual phase, through Ohmic dissipation and continued evaporation. \citep{2016ApJ...820...14Q} suggested several other avenues for investigation. Flare loop retraction following reconnection releases energy \citep{1996ApJ...459..330F}, which has been modelled using the \texttt{DEFT} and \texttt{PREFT} codes \citep{2010ApJ...718.1476G, 2015ApJ...813..131L}. If retracting loops receive resistance from lower-lying loops, slow shocks can be generated, resulting in heating over longer timescales \citep{1982SoPh...76..357C,1983ApJ...266..383C}. If loops do not reach their lowest energy state following impulsive energy release, they may continue to release energy via Joule dissipation, undergoing slow reconnection. Finally, the generation of magnetohydrodynamic (MHD) waves and their damping might last several tens of minutes \citep{2011SSRv..158..397W}. These processes should be modelled, and the energy released by each mechanism should be compared to the observed gradual-phase requirements. This will require going beyond 1D in hydrodynamic models, and will require 2D/3D flare MHD codes. 

To derive effective observational constraints on these models, we need unsaturated, high cadence, high spatial resolution images covering a broad temperature range of flare sources, measurements of loop length (and ideally loop geometry), and the timescale between successive loop activations. During the impulsive phase, chromospheric material that flows up into the corona (chromospheric ablation) can be diagnosed by spectroscopic observations. For example, the Doppler shifts of the Fe~\textsc{xxi} 1354.1~\AA\ line (forming at $\sim11$~MK) from thousands of pixels in a single flare were found to be remarkably consistent: the narrow distribution slowed from peak shift back to rest over several minutes \citep{2015ApJ...807L..22G}. When modelled, however, this Doppler shift persists only as long as energy is injected, so there is no gradual-phase Doppler shift within each pixel \citep{2020ApJ...900...18K}. The mechanism(s) by which continued energy deposition takes place must be able to reproduce such observations.

The presence of QPPs in the emission from flares provides crucial information regarding energy-release timescales and heating processes \citep{2009SSRv..149..119N}. While QPPs are most often prominently identified during the impulsive phase in emission associated with nonthermal particle acceleration, small-amplitude QPPs associated with hot thermal phase (e.g., SXR) can extend late into the gradual phase \citep{dennis2017detection,hayes2019persistent}, continuing for a significant period of time after the last detectable signatures of particle acceleration \citep{hayes2019persistent}. The underlying mechanism of QPPs is debated, but it is generally agreed that they are driven by either MHD waves or repetitive (or `bursty') reconnection \citep{2009SSRv..149..119N, 2018SSRv..214...45M}. Both mechanisms depend on properties of the flare reconnection region and the acceleration sites, and thus are directly associated with flare energy release processes. Following the recent identification of gradual-phase QPPs, many questions arise about the nature of these pulsations, their relation to their impulsive-phase counterparts, and whether they are signatures of extended energy release and heating, given that they are prominently identified in the hottest EUV channels and SXR observations. To use the observational signatures (amplitude and period) of gradual-phase QPPs to identify the energy release process, we must identify and locate the mechanism(s) responsible. These small-amplitude QPPs need to be spatially and temporally resolved in SXR and EUV images with high time cadence, on the order of seconds. Crucially in order to identify pixel locations of the QPPs, which occur in the brightest regions, these images must not be saturated.  An EUV instrument that is specifically flare-designed would be key, such that it has short exposure control that does not vary much over the duration of the flare.

\vspace{-0.15in}
\section{Suppressing Thermal Conduction}
\vspace{-0.15in}

Flare loops cool primarily through radiation and conduction. To properly assess the need for energy deposition in the gradual phase, we must ensure that important plasma physics effects are included in our models, and that detailed observations are used to estimate the magnitude of their impact. For example, the suppression of thermal conduction via nonlocal effects and turbulent scattering would slow the conductive cooling substantially, lengthening the gradual phase in simulations.

Classical Spitzer-H{\"a}rm conduction is essentially a collisional process. When the collisional mean free path (MFP) for the electrons responsible for carrying the heat flux is long compared to the scale of the temperature gradient, then non-local heat transport can be important \citep{1987ApJ...320..904K}. During solar flares steep temperature gradients routinely form through the corona and in shocks in the chromosphere. Turbulence is also present, leading to turbulent scattering and a further reduction of the heat flux \citep[][]{2017PhRvL.118o5101K,2018ApJ...865...67E}. Accounting for both effects \citep[][]{2016ApJ...824...78B,2016ApJ...833...76B,2017ApJ...835..262B,2018ApJ...852..127B,2018ApJ...865...67E} was found to produce a heat flux significantly smaller that than predicted by Spitzer-H{\"a}rm conduction. Experiments using the \texttt{EBTEL} code to model flaring coronal loops \citep{2018ApJ...852..127B} showed that reducing the heat flux resulted in cooling timescales more consistent with observations. Given the significant impacts that non-local effects and turbulent scattering will have on both the impulsive- and gradual-phase dynamics, these effects should be incorporated into state-of-the-art flare radiation-hydrodynamic models to properly account for non-local effects (and in MHD models that have thermodynamics). A convenient method to incorporate both non-local and turbulent effects was introduced recently by \citep{2018ApJ...865...67E}. Note that reducing the heat flux through the transition region into the chromosphere is also likely to yield longer duration mass flows. 

Numerical experiments that cover the relevant parameter space of turbulent MFP length and  spectral index should be performed, building upon the efforts of \citep{2018ApJ...852..127B}. The aim is to determine the most likely values of those parameters and the magnitude of suppression of conduction, with predicted peak temperatures and cooling times compared to observations of those quantities. An estimate of the turbulent mean free path length was calculated using observations of the variation of hard X-ray source size with energy \citep{2014ApJ...780..176K}, and the kinetic energy of turbulent motions (but not its spectrum) has been estimated using a combination of X-ray and EUV spectroscopy \citep{2017PhRvL.118o5101K}. Future observations should aim to determine both the turbulent MFP and the turbulence spectrum of fluctuations of the magnetic field present in flares, ideally simultaneously. These quantities should be measured from multiple heights spanning chromosphere to corona (thus spanning multiple temperature regimes), and will require simultaneous high spatial resolution hard X-ray observations and (E)UV (arcsecond or better), and high spectral resolution (E)UV spectroscopy (a few m\AA\ or better) to observe nonthermal line widths \citep[e.g.,][]{2018SciA....4.2794J}. Estimating the magnitude of suppression from observations is also possible \cite{2015ApJ...811L..13W}. There, slow-mode waves were observed in a flaring coronal loop, and the phase shift between perturbations of temperature and electron density was obtained. From the values measured, it was suggested that thermal conduction was reduced from the classical value, and that this effect grew with temperature. 

Through resolving the serious problem of the gradual-phase energy budget, we will also learn about turbulence in the solar atmosphere at a range of scales. Thus we can probe fundamental plasma physics processes at work in the Sun that will also be at play (on different scales) throughout the heliosphere, with far-reaching consequences beyond governing the flare gradual phase. Turbulence, for example, could play a role in the heating of the corona and acceleration of the solar wind \citep{cranmerreview}. The ubiquitous conversion of magnetic  and kinetic energy to heating in the solar atmosphere is undoubtedly impacted by turbulence. Non-local heat transport  also affects the formation of the transition region, and the flow of mass and energy between the chromosphere to corona.


\begin{spacing}{1}
\vspace{-0.1in}
\begin{multicols}{2}
	\bibliography{Helio2050_GSKerr}
\end{multicols}
\end{spacing}
\end{document}